\documentclass[aps,prb,twocolumn,showpacs,floatfix]{revtex4}
\usepackage{graphicx}

\begin{document}

\title{Effect of the lattice misfit on the equilibrium shape of strained 
islands in Volmer-Weber growth}
\author{Jos\'e Emilio Prieto}
\email{joseemilio.prieto@uam.es}
\affiliation{Centro de Microan\'alisis de Materiales (CMAM), \\
Dpto. F\'\i{}sica de la Materia Condensada and Instituto ``Nicol\'as Cabrera'', Universidad Aut\'onoma de Madrid, E-28049 Madrid, Spain}

\author{Ivan Markov}
\email{imarkov@ipc.bas.bg}
\affiliation{Institute of Physical Chemistry, Bulgarian Academy of Sciences,
1113 Sofia, Bulgaria}

\date{\today}

\begin{abstract}
We have studied the effect of the misfit on the equilibrium shape of
three-dimensional pyramidal islands grown on a foreign substrate in the case
of incomplete wetting (Volmer-Weber mode of growth).
By means of atomistic simulations using anharmonic interaction potentials, 
we find that tensile islands have smaller aspect ratios compared with 
compressed islands owing to their better adhesion to the substrate. The average
strains of consecutive layers decrease faster with thickness in compressed
than in tensile islands. The strains decrease rapidly with thickness, with
the consequence that above a certain height, the upper layers of the pyramid 
become practically unstrained and do not contribute to a further reduction 
in the top base. As a result, the truncated pyramids are not expected to 
transform into full pyramids. Our results are in good agreement with 
experimental observations in different systems. 
\end{abstract}


\maketitle

Elastic strains play an important role in the growth of heteroepitaxial thin
films. In the case of complete wetting of the substrate by the film material,
the el\-ast\-ic strain due to the lattice misfit results in a trans\-i\-tion
from planar to three-dimensional (3D) island growth beyond a critical 
thickness.\cite{Bau58,Pei78} This is the well known Stranski-Krastanov (SK) 
mode of growth, which is widely used to produce quantum dots in semiconductor 
systems such as Ge/Si\cite{Eag90} and InAs/GaAs\cite{Joy04}. In the case of 
incomplete wetting (non-zero wetting angle), 3D islanding takes place from the 
very beginning of the deposition, irrespective of the value of the mis\-fit. 
Nevertheless, in this case of Volmer-Weber (VW) growth, the misfit can play a 
crucial role in determining the equilibrium shape and, in particular, 
the aspect ratio of the 3D islands. The two cases mentioned above differ 
con\-si\-derably. In the SK mode, complete wetting and mis\-fit strain 
operate in different directions: complete wetting favours planar growth, 
while strain favours clustering.
This gives rise to an in\-sta\-bi\-lity of planar growth against 3D 
islanding. In the VW mode both the incomplete wetting and the effect of 
misfit strain favor islanding.

For the reasons given above, the equilibrium shape of islands on misfitting
substrates has been the object of intensive
studies\cite{Voor94,Spens97,Moll98,Daru97,Uemu02,Vill97,Shk06}. The
dependence of the equilibrium shape in both SK and VW modes has been
considered in detail by M\"uller and Kern.\cite{Mul96,Mul98,Mul00} In the
case of VW growth, they found that box-shaped crystals thicken with increasing
value of the misfit. A qualitatively similar conclusion was reached in 
Ref.~[\onlinecite{Mar95}] under the condition that the crystal does not relax
and is homogeneously strained 
to fit the substrate. In the case of pyramidal islands, the above authors  
found that the upper base shrinks and the island shape can change from 
truncated to complete pyramids with increasing misfit. It has been found
in addition that the equilibrium shape depends on the square of the lattice
misfit thus being independent of the misfit sign.\cite{Mul00} M\"uller
and Kern explained this result by neglecting surface stress effects
in their calculations.

\begin{figure}[t]
\includegraphics*[width=8.5cm]{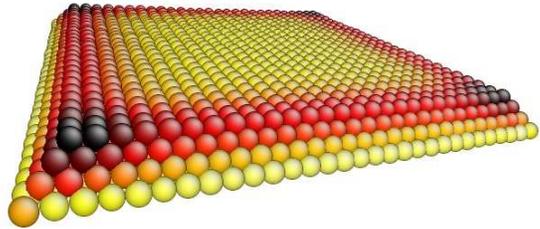}
\caption{\label{nicepict}
(Color online) Epitaxial fcc(100) island of
-7\% misfit on a rigid substrate. It has the shape
of a truncated square pyramid with 30$\times$30 atoms in the base plane and a
height of 4 monolayers. The color scale denotes the level of hydrostatic
strain as given by the atomic local strain tensor~\cite{Shimizu} and has
been represented using the AtomEye software~\cite{Li}. Strain is
highest at the center of the island and at the bottom layer and it is relaxed
at island edges and at the topmost layer.}
\end{figure}

In the present paper we study the effect of the lattice mis\-fit, and in 
particular, of its sign on the equi\-lib\-ri\-um shape of pyramidal crystals. 
We performed atomistic simulations considering square-based py\-ra\-midal 
islands with fcc atomic structure and (100) orientation, located on a 
substrate to which periodic boundary conditions are applied.
The equilibrium shape of fcc crystals considering only nearest neighbor 
interactions is a trun\-cat\-ed octahedron or a cube-octahedron enclosed
by six (001) and eight (111) faces with equal edge lenghts.\cite{Hen05} 
The wetting function (or adhesion parameter)
is defined as\cite{Kai50,Kai60}
\begin{eqnarray}\label{phi}
\phi = 1 - \frac{\psi^{\prime}}{\psi}
\end{eqnarray}
where $\psi ^{\prime} $ and $\psi $ are the substrate-deposit (adhesion)
and deposit-deposit (cohesion) atomic interaction energies, 
re\-s\-pec\-ti\-ve\-ly.
If the wetting function is smaller than 0.25 at zero misfit, the lateral cubic
faces do not appear and the equi\-lib\-rium shape remains a simple truncated
pyram\-id (Fig.\ \ref{nicepict}). The lattice misfit decreases the degree of
wetting and contributes to an effective increase in the wetting
function.\cite{Kor00,Pri02} For this reason, we restrict ourselves to values
of the wetting in the interval $\phi = 0 \div 0.2$.

We have performed simple minimization calculations. The atoms in the islands 
interact through
a pair potential whose anharmonicity can be varied by adjusting two constants
$\mu $ and $\nu $ ($\mu > \nu $),\cite{Mar93,Pri07}
\begin{equation}
V(r) = V_0\Bigl[\frac{\nu }{\mu - \nu}e^{-\mu (r-b)} - \frac{\mu}{\mu - \nu}
e^{-\nu(r-b)}\Bigr],
\end{equation}
where $b$ is the equilibrium atom separation. For $\mu = 2\nu$ the 
potential adopts the familiar Morse form. It is worth to note that 
these kinds of potentials, due to their spherical symmetry, are better 
suited to describe the bonding in metals. By contrary, semiconductors 
are characterized by directional, covalent bonds and bonding angles need 
to be taken into account. However, the potentials appropriate for 
semiconductors are also anharmonic~\cite{Ter88}, so we expect our results 
using anharmonic potentials to be qualitatively valid for those materials 
as well.

Energy minimization is performed by allowing the atoms to relax until the 
forces fall below some negligible cutoff value. In spite of its simplicity, 
the above potential includes the necessary features to describe real 
materials (the\-o\-re\-tical strength and anharmonicity) by varying the 
constants $\mu$, $\nu$ and $V_0$. 
We consider interaction only in the first coordination sphere and 
simulate both rigid and relaxable substrates, in which a given number 
of layers $S$ are allowed to relax. The initial positions of the atoms 
correspond to the centers of the potential troughs provided by their
neighbors underneath; those neighbors are not updated when atomic 
displacements approach values close to one half of the interatomic 
distance, so our model is not appropriate for the description of 
configurations containing misfit dislocations. 
The interatomic spacing in the substrate material is $a$ so that the lattice 
misfit is given by $f = (b-a)/a$. The case of positive misfits corresponds to 
the lattice parameter of the deposit being larger than that of the substrate, 
so the material in the islands is compressed. The atoms interact with 
their lateral neighbors 
mainly through the stronger repulsive branch of the potential. In the
opposite case, the islands are tensile and the atoms interact with their
lateral neighbors with the weaker attractive branch of the potential. As a
result, the surface stress of the side walls should differ in compressed and
tensile crystallites, or in other words, they should relax in different
degrees. The side walls of compressed islands are expected to relax by a 
greater amount compared with tensile islands.

In addition, the misfit sign affects the adhesion of the crystals to the
substrate in a different way. In compressed islands the atoms of the first
atomic layer are strongly displaced from the bottom of the corresponding
potential troughs of the substrate owing to the stronger repulsion of the
lateral neighbors. As a result, the adhesion is reduced. 
On the contrary, 
the first-layer atoms in tensile islands are only slightly displaced from 
the bottoms of the respective potential troughs of the substrate and the 
adhesion is stronger compared with compressed islands at the same absolute 
value of the lattice misfit\cite{Mar84a,Mar84b}.

\begin{figure}[h,t]
\includegraphics*[width=7.5cm]{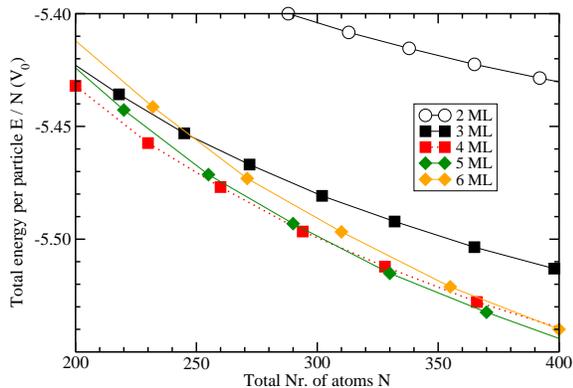}
\caption{\label{eqshape}
(Color online) Dependence of the total energy per atom as a function of
the number of atoms
$N$ in the pyramid for islands with different heights varying through one
monolayer. As seen islands with $N < 300$ have lowest energy when they are
four monolayers high whereas the lowest energy height is 5 monolayers at
$N > 300$. The wetting parameter is $\phi$ = 0.2, the lattice misfit~$f$ is
zero and a rigid substrate was assumed in the simulations.}
\end{figure}

In order to determine the equilibrium shape of the islands, we apply the 
following procedure. We consider islands of different heights and bases 
with shapes close to squares, either of $n \times n$ or of $n \times (n+1)$ 
atoms, where $n$ is an integer. Allowing the structures to relax, we 
calculate the total energy per atom 
of the islands, for heights $h$ varying by 1~monolayer (ML), as a function 
of the total number of atoms $N$ in the pyramid. Due to the discrete character
of the island sizes produced in this way, we need to interpolate between 
the two closest values at the corresponding height to the desired total 
number of atoms, e.g. $N$ = 1000.  In this way, we find the 
height of the island with lowest energy. For example, 
Fig.\ \ref{eqshape} shows that four-layers thick islands have lowest energy
when the number of atoms is smaller than 300. At $N > 300$ the lowest energy
thickness is 5 MLs. We then calculate the lowest energy heights of
islands at a given constant number of atoms and for different values of the
lattice misfit. Once we know the lowest energy heights we determine the edge
lengths of the lower, $n$, and upper, $n^{\prime}$, bases from the relation
$n^{\prime} = n - h + 1$ and from the formula for the total number of 
atoms $N(n,h)
= \sum_{k=1}^{n}k^2 - \sum_{k=1}^{n-h}k^2$ as a function of $n$ and
$h$. The height $h$ is measured in number of monolayers and $n$ and $n^{\prime}$
give the number of atoms in the upper and lower edges, respectively. 
Thus, the full pyramid is characterized by $n^{\prime} = 1$ and $h/n = 1$.

The simplest expression for the equilibrium shape of the pyramid is
$h/n^{\prime} = 4\phi$, where $\phi$ is the wetting function defined above
and the coefficient 4 reflects the coordination of an atom on the (001)
surface. This formula has been derived by comparing the works per atom 
to evaporate whole atomic planes of the lateral pyramidal faces and the
upper base, a method which had been introduced by Stranski and
Kaischew\cite{Str34} (an excellent recent review is given in 
Ref.~\onlinecite{Tas08}). However, we will illustrate our results 
in terms of plots of the ratio $h/n$ vs. the lattice misfit because 
the ratio $h/n^{\prime}$ tends to the island height $h$, which depends 
on the number $N$ of atoms in the pyramid when it becomes complete 
and the upper base disappears, while the ratio $h/n$ tends to unity.

\begin{figure}[h,t]
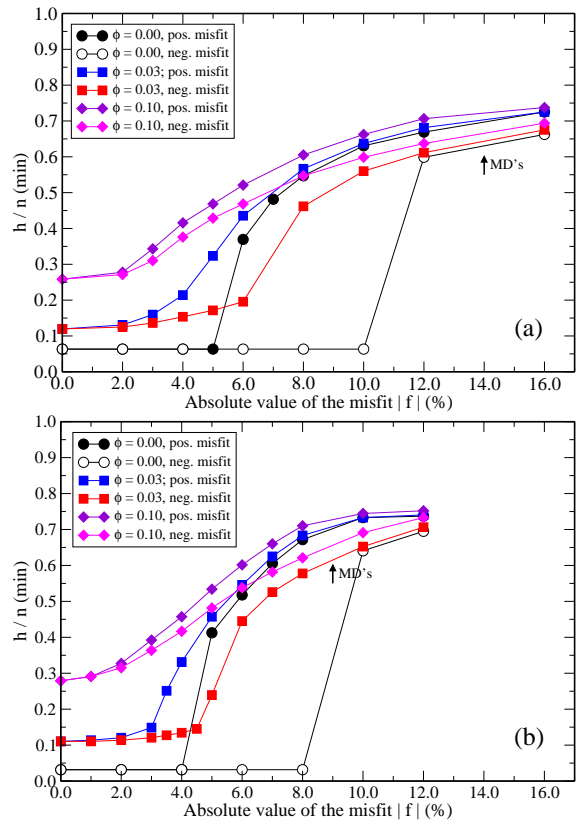

\includegraphics*[width=7.5cm]{./Fig3a.eps}
\includegraphics*[width=7.5cm]{./Fig3b.eps}
\caption{\label{asprat}
(Color online) Misfit dependence of the aspect ratio $h/n$ for different values
of the wetting function $\phi = 0.0, 0.03$ and 0.10 and a total number of atoms
in the pyramid of (a) $N = 250$, (b) $N = 1000$. A rigid substrate was assumed
in the calculations. The arrows mark the appearance of misfit dislocations in the 
growing islands.}
\end{figure}

Figure\ \ref{asprat} demonstrates our main result. It shows the dependence of
the aspect ratio $h/n$ of islands of mi\-ni\-mum energy on the value of the 
misfit, both negative and positive ones. The islands consist of small (a) 
$N = 250$ and large (b) $N = 1000$ number of atoms. The substrate is assumed 
rigid. Several important conclusions can be reached. First, the lattice
misfit has a great effect on the equilibrium shape. The aspect ratio
increases steeply beyond a moderate misfit of about 3\%. The smaller the
wetting function (i.e. the stronger the adhesion), the stronger the effect of
the misfit. Or, the stronger the tendency to 3D islanding due to the weaker
wetting, the smaller is the effect of the lattice misfit. The effect of the
misfit is greatest at $\phi \rightarrow 0$. As mentioned above, the reason is
that both effects are complementary. 

Second, it is seen that a positive misfit has a per\-cep\-tibly greater effect 
on the equilibrium shape than a negative one of the same absolute value.
This is due to the stronger repulsive forces of the inter\-atomic potential
and to the weaker ad\-he\-sion. In ten\-sile over\-growth the increase in the
aspect ratio is de\-layed to greater absolute values of the misfit. And more
important, increasing the lattice misfit does not lead to disappearance of 
the upper base, $h/n \rightarrow 1$, as predicted by M\"uller and
Kern\cite{Mul00}. In both cases of positive and negative misfit the aspect
ratio $h/n$ goes to a saturation value of about 0.75 irrespective of the 
island size and misfit sign.

The curves $h/n$ vs misfit show an initial increase and an inflection point. 
Beyond this point the curves tend to a saturation value lower than unity
as mentioned above. Note that at large absolute values of the misfit the 
in\-ter\-face between the crystallites and the substrate is expected to 
be resolved in a cross grid of misfit dislocations. The arrows in 
Fig.\ \ref{asprat} show the approximate values of the misfit at which
dislocations are introduced in compressed islands. The tensile islands, on
the other hand remain coherent. Although our model, as explained above, is not
well suited for the description of dislocated configurations, it seems clear
that the tendency to saturation begins long before the introduction of 
dislocations. Furthermore, since both 3D clustering as well as the 
introduction of misfit dislocations are (possible competing) mechanisms 
that contribute to the 
relaxation of epitaxial strain, we can conclude that the introduction of 
misfit dislocations is not the basic reason for the islands to remain 
truncated pyramids at the sizes studied so far. As will be shown below 
the main reason is the fast decrease in the strain with the pyramid height. 

We can understand the different behavior of the aspect ratio at positive 
and negative misfits as follows. As shown in 1+1 dimensional 
models\cite{Kor00,Mar02}, the thermodynamic driving force for 3D islanding 
in coherent Stranski-Krastanov growth is the reduced adhesion of the initial 
2D islands to the wetting layer. This is due to the
displacements of the atoms located closer to the islands edges from the
respective potential troughs of the substrate (the wetting layer). Thus the
average adhesion parameter is non-zero and 3D islanding is
thermodynamically favored. A more detailed study of the adhesion parameter
as a function of the misfit sign has shown that compressed islands display
larger values of the adhesion parameter (weaker wetting) compared with
tensile islands. Hence, greater absolute values of the (negative) misfit
in tensile islands are required in order to reach the same values of the
adhesion parameter as compared to compressed islands.

\begin{figure}[h,t]
\includegraphics*[width=7.5cm]{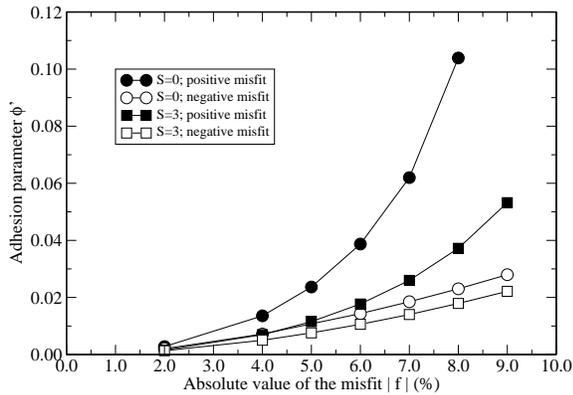}
\caption{\label{adhesion}
Misfit dependence of the adhesion parameter, $\phi ^{\prime}(f)$, for
compressed and tensile islands for the case of $\phi = 0.0$ (SK mode of growth).
Results are shown for different numbers~S of substrate layers allowed to relax:
circles for S~=~0 (i.e., rigid substrate) and squares for S~=~3~ML. The
islands are 5~ML high and contain a total amount of $N = 990$ atoms.}
\end{figure}

The same result is obtained in our 2+1 dimensional model as demonstrated
in Fig.\ \ref{adhesion} for the Stranski-Krastanov growth. 
Here the wetting parameter $\phi$ due to different bonding, as defined 
by Eq.(\ref{phi}), equals zero
and we can define a mean wetting parameter $\phi^{\prime}$ as
\begin{equation}
\phi^{\prime} = 1 - \frac{U_{\rm adh}}{(- 4 V_0)},
\end{equation}
where $U_{\rm adh}$ is the mean interaction energy between an atom at 
the bottom layer of the island and the substrate, which equals -4$V_0$ 
at zero misfit. The tensile is\-lands always display better adhesion 
to the substrate. Although the difference in adhesion is greater in 
rigid substrates compared with relaxed ones, the effect of the 
potential anharmonicity remains significant. The adhesion pa\-ra\-met\-er 
for tensile islands remains close to zero (com\-plete wetting).

In the case of Volmer-Weber growth the effective adhesion parameter in
misfitting overgrowth, $\phi_{\rm eff}$, is thus composed of two parts. The
first part is due to the difference in bonding and is given by Eq.\ (\ref{phi}),
while the second part is due to the lattice misfit as discussed above. We can
thus write
\begin{equation}
\phi_{\rm eff}(f) = \phi + \phi^{\prime}(f)
\end{equation}
where the second term in the right-hand side depends solely on the lattice
misfit, $f$, so that $\phi^{\prime}(0) = 0$. The inequality 
$\phi^{\prime}(+|f|)>\phi^{\prime}(-|f|)$ and in turn 
$\phi_{\rm eff}(+|f|)>\phi_{\rm eff}(-|f|)$
explains the results shown in Fig.\ \ref{asprat}.

\begin{figure}[h,t]
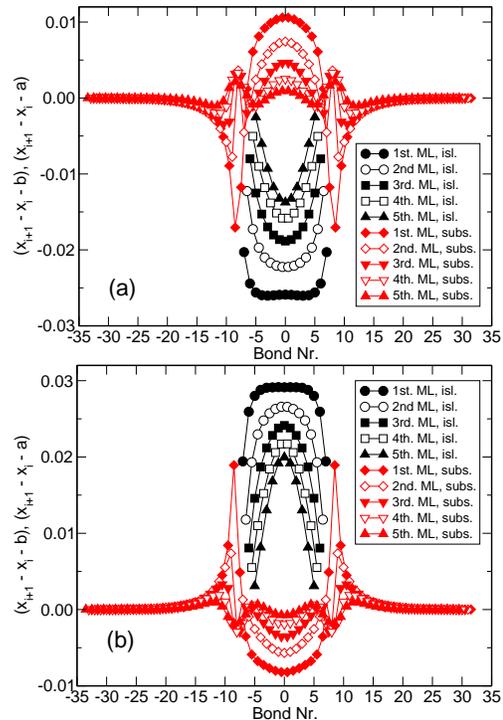

\includegraphics*[width=6.5cm]{./Fig5a.eps}
\includegraphics*[width=6.5cm]{./Fig5b.eps}
\caption{\label{strains}
(Color online) Distribution of the lateral bond strains in the center of the
crystal and in the
layers of the substrate underneath for misfits of (a) f~=~4\% and (b) f~=~-4\%.
Strains are referred to the lattice parameters of crystal $a$
and substrate $b$, respectively. The wetting parameter due to the different
chemical nature of substrate and adsorbate is $\phi = 0.03$, islands of
5~ML height and 990 atoms were considered and 5~ML of the substrate were
allowed to relax.}
\end{figure}

We now study the
strain distribution in tensile and compressed islands on a substrate which is
allowed to relax. Fig.\ \ref{strains}(a) shows the distribution of the strains
in a compressed crystallite ($f = +4\%$). The crystal is 5~MLs thick and the
same number of monolayers of the substrate are allowed to relax. As seen, the
4th and the 5th MLs of the substrate are very weakly strained. The strains in
the substrate below the island do not exceed 1\% whereas those in the first 
layer of the crystallite are close to -3\%. 

Figure\ \ref{strains}(b) shows the corresponding strain distribution in 
tensile island at the same absolute value of the misfit ($f = - 4\%$). 
The distribution is qualitatively the same. The strains in the island are 
again much larger than those in the substrate. This is easy to understand 
having in mind that the
substrate crystal has no free edges and the substrate atoms are highly
constrained by their lateral neighbors. As a result, the total strain energy
stored in both compressed and tensile islands is about 
two orders of magnitude larger than the strain energy stored in the substrate. 
The important conclusion we can extract is that the substrate behaves as 
relatively ``stiff" against the epitaxial strain induced by the 
overgrowth and can be safely considered rigid in comparison with the 
deformations of the islands. This justifies the results in 
Fig.\ \ref{asprat} where the substrate was assumed to be rigid.

\begin{figure}[h,t]
\includegraphics*[width=6.5cm]{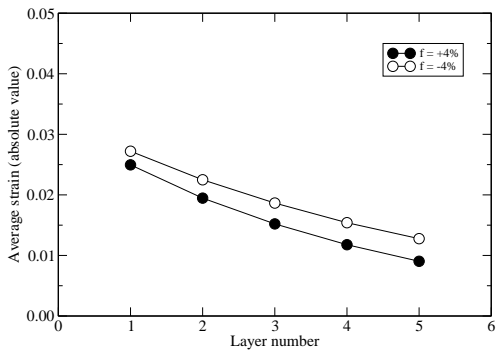}
\caption{\label{constrain}
Average strains of consecutive layers in compressed and tensile pyramids with
+4.0\% and -4.0\% lattice misfit, respectively. Islands are 5~ML high and contain
990 atoms, the wetting parameter is $\phi$ = 0.03 and 5~ML of the substrate were
allowed to relax.}
\end{figure}

A more careful inspection of Fig.\ \ref{strains} shows that the lateral
strains of the separate layers in compressed islands are always smaller than 
the corresponding strains in tensile islands. In addition, the strains of the
separate layers decrease with layer height, this decrease being larger 
in compressed than in tensile islands (Fig.\ \ref{constrain}).
In other words, compressed islands tend faster to achieve the bulk lattice
parameter of the overgrowth material compared with tensile islands.

\begin{figure}[h,t]
\includegraphics*[width=6.5cm]{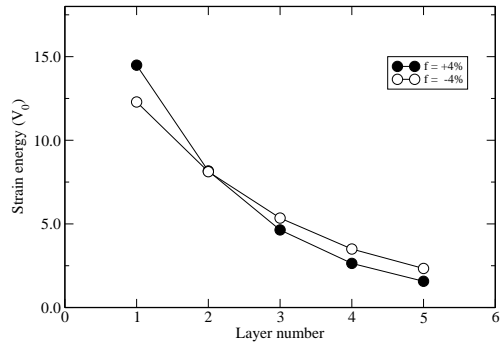}
\caption{\label{estrheight}
Strain energy of consecutive layers in compressed and tensile pyramids with
+4.0\% and -4.0\% lattice misfit, respectively. Islands are 5~ML high and contain
990 atoms, the wetting parameter is $\phi$ = 0.03 and 3~ML of the substrate were
allowed to relax.}
\end{figure}

However, the
height distribution of the strain energies of the separate layers shows an
unexpected behavior (Fig.\ \ref{estrheight}). The strain energies of the
layers beyond the second one in tensile islands become larger than those
in compressed layers. This can be understood as follows. The more
strained the layer the more significant is the effect of the po\-ten\-tial
anharmonicity. The first layers are highly strain\-ed and the steeper
repulsive branch of the potential leads to a very high strain energy of the
compressed island. The tendency of atoms to relax the strain is thus higher
in compressed overlayers and this leads to smaller values of the strain 
in spite of the higher amount of strain energy as compared to tensile islands
at the same absolute value of the lattice misfit. Upper layers become 
less and less strained and the anharmonicity cannot overcompensate 
the smaller absolute values of the strains. As a result, the more strained 
upper layers in tensile islands possess also a larger amount of strain energy.

Figure\ \ref{strains} also shows that the first layer is practically 
pseudomorphic with the substrate (the strains are nearly equal to the 
misfit taken with opposite sign) with the exception of a few edge bonds.
On the contrary, most of the bonds in the uppermost (fifth) layer are much 
less strained than the nominal misfit showing that the strains rapidly 
goes to zero. 
It follows that the upper layers of a complete pyramid are practically 
fully relaxed and so they do not contribute to the accumulation of strain in 
the crystallite. 
For this reason, the crystallites do not transform to full pyramids for
reasonable values of the misfit as expected from the continuous and harmonic 
theory of M\"uller and Kern. The presence of misfit dislocations favors 
further relaxation of the upper layes, but they are expected to play a 
secondary role in preventing the transformation of the small islands into 
full pyramids. An additional argument is that misfit dislocations are 
introduced in tensile islands always at misfit values where saturation 
takes place. In order to further clarify the role of the anharmonicity of the
potential, we also studied the island's aspect ratio as a function of the
lattice misfit using harmonic potentials of the same spring constants 
as the anharmonic ones previously discussed. The curves (not shown) 
obtained for positive and negative misfits nearly overlap and are located 
as expected between the curves corresponding to the anharmonic potentials, 
slightly closer to the curve of positive misfit.


It is worth to note that the equilibrium aspect ratio of the 3D islands in the
case of complete wetting (SK mode) follows the same trend as in the case of
VW growth mode (Fig.\ \ref{asprat}). The only difference is that the initial
islands are single-layer islands ($\phi = 0$) which then transform into
multilayer islands\cite{Pri05}. The only physical reason for incomplete
wetting is the lattice misfit which leads to displacements of the overlayer
atoms from the bottom of the substrate potential troughs\cite{Kor00,Pri02}
and thus to a weakening of the interfacial bonding.

Experimental observations of equilibrium shape of islands on misfitting 
substrates, in the appropriate ranges of parameters to compare with the results
of our model, are scarce in the literature. We will not discuss here the 
results concerning the equilibrium shape of metal clusters on MgO 
(for a review see Ref. \onlinecite{Hen05}). One of the reasons is that 
wetting functions are greater than 0.5 and so the equilibrium shapes are not 
simple pyramids. Another reason is that all metals studied (Ni, Pd, Pt) 
show a negative misfit with the substrate\cite{Mot07}. Nevertheless, smallest
particles of Pd on MgO (f = -6\%) studied showed an
equilibrium shape of a truncated pyramid, $h/n \approx 0.7$\cite{Gio90}.

Gold\-farb et al.\cite{Gold05} found that the
equilibrium shape of small crystals of TiSi$_2$ on Si(111) is a flat
hexagonal pyramid with the ($01\bar3$) atomic plane parallel to Si(111).
The misfit in the two directions $<112>$ and $<110>$ amounts to 5\% and 10\%,
respectively. Silly and Castell studied a completely different system: Fe on
SrTiO$_3$(001)\cite{Sill05}. After annealing, the equilibrium shape of 
bcc-Fe islands is a square pyramid with an aspect ratio $l/h = 1.2$ 
irrespective of the island size, where $l$ is the edge length of the upper 
base. The side walls make an angle of $45^{\circ}$ with the substrate so 
that the aspect ratio $h/L$ amounts to 0.31, where $L$ is the
edge length of the lower base. Nowicki et al. found that the equilibrium
shape of Pb islands on Ru(001) corresponds to a truncated pyramid with the 
(111) face as the upper base\cite{Now02,Bon07}. So in all cases discussed 
above, the equilibrium shape is a truncated pyramid in accordance with 
the results of the present paper.

In conclusion, using anharmonic interaction potentials, we have found 
that compressed islands show larger values of the aspect ratio than 
tensile ones at the same absolute value of the misfit owing to
its larger effective adhesion parameter. Atoms belonging to the first
layer of compressed islands show larger displacements from the bottom of the
potential troughs of the substrate due to the stronger interatomic repulsive 
forces, which leads to weaker adhesion of the islands to the substrate. On
the contrary, the weaker attractive branch of the potential acts between the
atoms in tensile islands and the effective wetting parameter is smaller. Due
to the same reasons, compressed islands are less strained compared with
tensile ones and tend to attend the bulk lattice parameter more rapidly with
increasing island thickness. As a result, the upper layers of the pyramids are
practically unstrained and they cannot transform to full pyramids with 
increasing values of the lattice misfit. The strain energy
stored in the substrate is about two orders of magnitude smaller than in the
islands due to the constraint of the atoms by their lateral neighbors. As a
result, for many calculations, the substrate can usually by assumed as rigid.

\acknowledgments This work was financed by project FIS2008-01431 of the 
Spanish MiCInn; J.E. Prieto also acknowledges support by its program 
``Ram\'on y Cajal".

\end{document}